\documentclass[runningheads]{llncs}
\usepackage{graphicx}
\usepackage{booktabs}
\usepackage{tabularx}
\usepackage{hyperref}
\begin{document}
\title{Monitoring the adoption of SPI-related \\best practices. An experience report\thanks{This work is part of a project that has received funding from the European Union's Horizon 2020 research and innovation programme under Grant Agreement No. 856726 (GN4-3). \protect\\The scientific/academic work is financed from financial resources for science in the
years 2019-2022 granted for the realisation of the international project co-financed
by the Polish Ministry of Science and Higher Education.}}
\titlerunning{Monitoring the adoption of SPI-related best practices. An experience report}
%

\author{Bartosz Walter\inst{1,5} 
\and
Branko Marović\inst{2} 
\and
Ivan Garnizov\inst{3}
\and
Marcin Wolski\inst{1}
\and
Andrijana Todosijevic\inst{4}
}
\authorrunning{B. Walter, B. Marović, I. Garnizov, M. Wolski, and A. Todosijevic}
%
\institute{PSNC, Poznań, Poland
\email{\{bartek.walter,marcin.wolski\}@man.poznan.pl}
\and
University of Belgrade, Belgrade, Serbia,
\email{branko.marovic@rcub.bg.ac.rs}
\and
Friedrich-Alexander-University, 
Erlangen, Germany,
\email{ivan.garnizov@fau.de}
\and
AMRES, Belgrade, Serbia,
\email{andrijana.todosijevic@amres.ac.rs}
\and
Poznań University of Technology, Poznań, Poland
}
\maketitle              
\begin{abstract}
Software Process Improvement requires significant effort related not only to the identification of relevant issues and providing an adequate response to them but also to the implementation and adoption of the changes. Best practices provide recommendations to software teams on how to address the identified objectives in practice, based on aggregated experience and knowledge. In the paper, we present the G\'{E}ANT experience and observations from the process of adopting the best practices, and present the setting we have been using.

\end{abstract}
\section{Introduction}
\label{sec:intro}



Software Process Improvement (SPI) is focused on finding more efficient methods of software development and optimizing the entire software life-cycle, to find an acceptable balance between expectations of stakeholders, constraints concerning cost and quality, and the capabilities of the development team. Until now, various approaches related to the design, implementation and evaluation of SPI have been identified, studied and verified~\cite{Unterkalmsteiner2014}. As a result, several different solutions have been proposed, implemented and are in use in the industry.

A typical SPI process includes several  activities or stages: it starts with capturing the problem, diagnosing it, finding a possible solution, and then applying it~\cite{Aysolmaz2011}. In many cases, the process is not linear nor sequential, but iterative~\cite{Stojanov2016}, resulting in a continuous and ever-improving procedure driven by collected feedback data.

In previous papers~\cite{Stanisavljevic2018,eurospi2019,Walter2020euro} we presented earlier stages of developing an SPI process in G\'{E}ANT, a pan-European project that originated from networking, but it currently includes also an important component of network-based services supported by software-intensive systems. G\'{E}ANT software teams are diverse in terms of language, nationality and work culture. In addition, they enjoy a high degree of independence, which also includes the freedom to choose, adapt and follow a specific approach to software development. This diversity creates a hospitable environment for experimenting and promoting innovations, but also provides a challenge to the coordination of work and managing the outcomes with respect to the expected deliverables and existing constraints.

In this work, we present the approach to the implementation, adoption and monitoring of best practices, aimed at optimizing and coordinating the software development process, and accompanied by preliminary observations of the process.

The paper is structured as follows. In Sec.~\ref{sec:relwork} we make a brief overview of the literature concerning the SPI, its implementation and support from the best practices. Sec.~\ref{sec:context} provides more details about G\'{E}ANT and its specific constraints that guide the entire SPI process. In Sec.~\ref{sec:monitoring} we describe  our approach to monitoring, based on a lightweight process aimed to support the teams in their efforts, but also providing valuable input to managers, in particular in the area of Product Lifecycle Management (PLM). In Sec.~\ref{sec:example} we present an exemplary implementation of some practices in one of the subject teams, and in Sec.~\ref{sec:summary} we summarize the work.

\section{Related work}
\label{sec:relwork}
Many aspects concerning the deployment and implementation of SPI initiatives have been discussed in the literature. In this section, we refer to selected works in that area.

As indicated by Kham and Keung, different standards and models for SPI have been developed to help organisations to improve their software processes and also numerous instruments are developed to evaluate process efﬁciency~\cite{Kham2016}. Moreover, they also emphasised the role of practice-originated recommendations in developing quality software systems. In some of the models, a large part of the SPI success is achieved through the implementation of best practices~\cite{Bayona2019}. It has been also discussed by Rahman et al.~\cite{Rahman2011}, who formulated best practices for ensuring quality in software engineering projects. Moreover, various patterns for the successful adoption of SPI practices have been discussed in the literature~\cite{singer2013improving}.


Proper monitoring of processes is another indispensable step in SPI.
Stojanov et al. proposed an approach based on frequent feedback, which enables reacting to deviations in short time~\cite{Stojanov2016}. 

Software process assessment, as the most important phase in process improvement, can be easily narrowed down to the list of processes that should be improved. Zarour et al. presented a systematic literature review aimed at investigating the best practices for the successful design and implementation of lightweight software process assessment methods~\cite{Zarour2015}. 

The role of a human factor in SPI efforts has been also widely recognized. The SPI Manifesto identifies key activities that contribute to successful SPI implementation. More than two-thirds of its recommendations refer to human engagement and initiatives. Messnarz et al.~\cite{Messnarz2014} discussed that issue also from the social responsibility perspective, introduced in the ISO 26000:2010 standard.

\section{Context}
\label{sec:context}


\noindent G\'{E}ANT software teams work in a specific environment that strongly affects the way they work. To a large extent, it results from a federated structure of the organization and the emphasis put on producing innovation by designing and developing prototypical applications based on the high-performance network, to support novel services for European researchers and citizens~\cite{Stanisavljevic2018}.

Teams in G\'{E}ANT share a common Product Life-cycle (PLC) and use common tools, but they can freely choose and customize their processes and techniques. Team members usually come from various national organizations (called NRENs), can be simultaneously involved in several projects, are geographically distributed, and have diverse cultural and professional backgrounds. This adds an additional complexity that needs to be addressed and embraced.

In response to that, work on Common Best Practices (CBPs)\footnote[1]{\url{https://wiki.geant.org/display/GSD/Catalogue+of+best+practices}} and G\'{E}ANT Software Maturity Model (SMM)\footnote[2]{\url{https://wiki.geant.org/display/GSD/Software+Maturity+Model}} has started in 2018. The preliminary versions of both frameworks have been reviewed by selected software teams, which resulted in revisions that are currently implemented in practice~\cite{Stanisavljevic2018,Walter2020euro}.

The CBPs for G\'{E}ANT software teams provide directional guidance and recommendations on how to achieve specific goals defined in the SMM~\cite{eurospi2019}, but they need to be customized and adapted by the teams to the local context. As a result, the generality of recommendations included in a practice appears to be a balanced trade-off between delivering actionable procedures on one hand and ensuring the necessary generality on the other~\cite{Walter2020euro}. 
A catalogue of Software Best Practices presented in Table~\ref{tab:cbp} covers all relevant aspects of software development that have been identified in G\'{E}ANT in response to the recognized needs and objectives~\cite{Wolski2017} and is one of the tools contributing to software governance. Currently, it includes 24 practices, and a few more are waiting for inclusion into the catalogue. The software governance team (SwM), formed as a part of one of G\'{E}ANT work packages, supports the teams in implementing and adjusting best practices. It also encourages them to share their experience and observations with other teams, creating a continuous improvement loop.

\begin{table}[ht]
\caption{Catalogue of Common Best Practices used in G\'{E}ANT. A--Requirements, B--Design \& implementation, C--Quality assurance, D--Team organization, E--Maintenance}
\begin{tabularx}{\linewidth}{p{0.1\linewidth} p{0.9\linewidth}}
\toprule
ID      & Practice                                                                    \\ 
\midrule
BP-A.1  & Identify an initial group of stakeholders and iteratively refine it \\
BP-A.2  & Elaborate a communication strategy for stakeholders                 \\
BP-A.3  & Identify, analyse and manage requirements                           \\
\midrule
BP-B.1  & Monitor and assess available technologies                           \\
BP-B.2  & Elaborate an approach for documenting processes and artifacts       \\
BP-B.3  & Track the changes in relevant artifacts                             \\
BP-B.4  & Support the build and delivery process with automation                  \\
BP-B.5  & Record and manage issues encountered with the product               \\
BP-B.6  & Verify licenses of the used libraries                               \\
\midrule
BP-C.1  & Identify and manage risk factors that could negatively affect the product \\
BP-C.2  & Identify relevant quality characteristics and test conditions and provide verification criteria for them \\
BP-C.3  & Elaborate and maintain a quality plan for the project               \\
BP-C.4  & Apply adequate methods to verify the outcome of development activities  \\
BP-C.5  & Monitor relevant quality parameters                                 \\
BP-C.6  & Periodically validate the product with project stakeholders         \\
BP-C.7  & Refine the quality assurance process based on retrospection         \\
\midrule
BP-D.1  & Manage skills of the team members and skills required in the project \\
BP-D.2  & Provide effective, adequate and reliable means for communication within the team \\
BP-D.3  & Define and implement an effective decision-making process           \\
BP-D.4  & Manage the schedules and assignments of the team members            \\
\midrule
BP-E.1  & Plan early and implement adequate maintainability-related activities  \\
BP-E.2  & Elaborate a strategy for managing maintenance issues                \\
BP-E.3  & Identify and evaluate available options for implementing a change request \\
BP-E.4  & Define a procedure for deploying changes to running services \\ 
\bottomrule
\end{tabularx}
\label{tab:cbp}
\end{table}

\section{Adoption and monitoring of best practices}
\label{sec:monitoring}



The process of implementing best practices and transferring them between various teams is organized into four iteratively repeated phases:
\begin{enumerate}
    \item Understanding the critical and important processes and practices in each team's work (by interviewing the team members),
    \item Reaching the agreement with the team concerning the implementation of SPI recommendations, making necessary adaptations and adjustments,
    \item Monitoring the adoption process and refining the approach,
    \item Evaluating the results, providing feedback and formulating new or updating the existing SPI-related recommendations.
\end{enumerate}

In the process, we intentionally avoid using the term "evaluation", which is associated with feedback not exclusively related to process improvement, but also the inevitable comparison of teams. Initial attempts to include an element of assessment was found to be too offensive by many teams and significantly hampered their motivation to participate in the SPI program. As a result, we decided to re-focus our attention on activities directly linked with process optimization.

In the following subsections, we provide more insights into key aspects of the process.


\subsection{Adaptation and operationalization of best practices}
\label{ssec:adaptation}
Adaptation is an indispensable step in the SPI process in the G\'{E}ANT context. Best practices that have been identified and documented do not deliver directly executable instructions, but directional recommendations at a high level of abstractness. This was a conscious decision, given the diversity of the solutions that addressed each objective in SMM, which could not be embraced by a single, universal practice. Consequently, the recommendations included in CBPs need to be customized and adapted to the local context of the implementing team. In practice, this happens in a sequence of interactions between SwM and the software team, with the following outline:
\begin{enumerate}
    \item Presentation of the catalogue of CBPs to the software team, with a concluding discussion;
    \item Selection of the practices, for which there exists the largest room for improvement;
    \item Presentation of the current status of the given objective;
    \item Adaptation and operationalization of the best practice, with an objective to address the specific context of the team;
    \item Defining preliminary metrics and parameters for monitoring the adoption process.
\end{enumerate}

In this process, two items deserve a comment. First, operationalization of the practice mostly concerns the translation of the generic recommendations of the practice into actionable statements that encompass the specific context of the team, its tools and procedures. Additionally, sometimes a specific practice also needs to be extended or narrowed, as some elements of it are out of scope in the team's context. As a result, the team can elaborate its own version of the practices, as long as their objectives are properly addressed.

The process of adoption needs to be controlled and managed based on quantitative values. Since \emph{modus operandi} varies among teams, the monitoring metrics are frequently identified in each case separately. As a side effect, due to the differences in metrics and parameters, it is virtually impossible to perform a cross-project comparative evaluation. However, we need to outline that the evaluating aspect of best practices adoption is not the top priority. 
At this stage, we only make a preliminary identification of the metrics, while they are fixed, tuned and calibrated in the next step. 

\subsection{Monitoring}
\label{ssec:monitoring}

Monitoring is another step in the deployment step that allows for controlling the process, collecting feedback and taking necessary corrective actions. It fosters the adoption of best practices, contributes to increasing the outreach of change, supports the consolidation of achieved results and is also a promotional tool that encourages the implementation of similar ventures in other teams.

The monitoring activity needs to be backed and guided by data, either quantitative or qualitative. However, since the initial status of software teams is diverse, and their specific needs and methods differ and are subject to subjective assessment and interpretation, it is very difficult to provide a single measure that would be universally applicable. 
The measurement can be also used as a feedback tool that will allow not only to evaluate the outcomes but also to validate and adjust them. 

\subsection{Tool support}
\label{ssec:tool}

To effectively support the monitoring process, adequate instrumentation is needed. The process of monitoring requires collecting  feedback data for tuning and taking corrective actions regularly. The feedback loop is similar to the one used for agile software development with two differences: it should require a lower level of engagement from the team (as the SPI process is usually only accompanying the primary activity), and it spans over a longer time, as it is not related to direct results of software development, but to the changes in development methods. 

To properly support the monitoring of the adoption of CBPs, we added a data collection and visualization plugin into the G\'{E}ANT Software Catalogue (GSC). GSC is a system that collects and manages information about all software projects in G\'{E}ANT. It is available to the SwM team, software teams and other stakeholders so that it provides continual access to all measurements and data. GSC regularly collects information from various components of the toolchain that  may be useful for this purpose, e.g., issue tracker, directory service or code repositories. 

The CBPs plugin relies on the data available in GSC but it also allows for defining  manually maintained metrics related to the adoption of CBPs in specific projects. In particular, the tool allows for defining arbitrary metrics that correspond to the process and entering the measurements by authorized stakeholders from G\'{E}ANT. For each metric we defined an ordinal scale of values; in many cases, a five-point Likert scale is used, which provides some uniformity across teams and practices. Stakeholders record changes in the monitored quantities and can observe the changes in the reported values.

In addition to the main metrics for individual best practices, it is also useful to monitor and discuss the related measures that quantify workflow results, such as source code metrics, incident frequency, number of served requests, user satisfaction, responsiveness or level of user participation, user retention and the like. They are also supported by the CBPs plugin and are used as customisable supporting and indicative metrics for individual best practices that provide evidence for the best practices or can be used to help in assisting its current status. These metrics are currently also manually collected, but in the future could be linked to the information that GSC already regularly collects from some of the tools it is integrated with.

Although all these quantities are currently monitored manually, in the future they could be extracted and analysed by an external tool. However, we do not expect that the monitoring will ever be fully automated, as the final appraisal of practice is highly context-dependent. 

\section{Example}
\label{sec:example}

\subsection{Adoption of three practices by the GSC team}

A few software teams started adopting selected practices that address the issues they currently face. One of the projects, codenamed GSC, is responsible for G\'{E}ANT Software Catalogue, a web system with data about software projects run by G\'{E}ANT. The team is small (3-4 people), started working in 2018 and applies Scrum-like, iterative approach to software development. 

After the initial discussion, the team decided to implement the following three practices from the catalogue that have the strongest impact on the team's performance.

\vspace*{1em}
\noindent\textbf{BP-A.1: Identify an initial group of stakeholders and iteratively refine it}

It is important, in particular at the initiation of the project, to identify and onboard all parties that could contribute their requirements, experience or feedback to the project. Innovative projects that lack appropriate support from their stakeholders usually miss their targets or deliver wrong products that meet only some of the requirements. The GSC team decided early to create and maintain a plain registry of individuals or teams related to the project. The registry is occasionally updated by removing inactive ones who are unlikely to restore their activity or who left their teams. According to the GSC team, the simplicity of this solution is a fair trade-off between the cost related to maintaining the registry and the advantage of having the core stakeholders identified. The registry is frequently used for consulting them about the next directions for the project or notifying them about releases and decisions.
Due to the projects' specifics and a relatively large number of stakeholders, the GSC team considers extracting the core ones who actively contribute to the project, to a separate list.

\vspace*{1em}
\noindent\textbf{BP-C.4: Apply adequate methods to verify the outcome of development activities}

This practice recommends the involved teams to balance the complexity and effort related to QA methods with the expected and actual outcomes. It does not mandate adopting specific testing or review methods, so it is quite generic. In response to that, the GSC team adopted three levels of testing that appeared highly effective in detecting defects and improving the identified quality characteristics: unit tests, integration tests and UI tests. They address the key concerns about the defects at the level of individual components, modules and the user-accessible interface. Currently, the tests are partially automated in areas where the effort vs. benefit ratio is the most promising. Additionally, the team is committed to improving the process during retrospection sessions.

\vspace*{1em}
\noindent\textbf{BP-E.4: Define a procedure for deploying changes to running services}

The process of releasing the product or its new versions may pose hazards for the users or other services. Therefore, it is essential to identify the key steps in that process, verify them and ensure they are consistently applied, for example in the form of a checklist. The GSC team adopted this procedure by identifying the parts which could be automated and the ones that need to be performed manually. The former ones have been defined by scripts, which backup the databases, re-generate configuration files or tag the code in the repository. The manual procedures embrace contacting the relevant service owners, who could be affected by the changes, and making sure that there are no other works planned in the infrastructure that could interfere with the release. Software Catalogue is one of the key services, and its high availability and stability is a core concern, which mandates assigning a high priority to the integrity of the release process. The entire procedure, initially scattered, is currently centralized, periodically reviewed by the team and maintained by the product owner and team manager. 
 
\subsection{Supporting the PLM with best practices} 
\label{ssec:plm}
Best practices are primarily oriented towards the software teams and focused on delivering the value directly to them. However, a similar need has been also recognized by the G\'{E}ANT management team as the body overseeing the IT service management and the Product Lifecycle Management (PLM) process. The collaboration between SMT and the management team resulted in applying the best practices framework to identify the activities that facilitate the management decisions on product and service maturity, but also support software development teams in meeting the objectives and standards defined by the management team.

\begin{figure}[h]
\includegraphics[width=12.2cm]{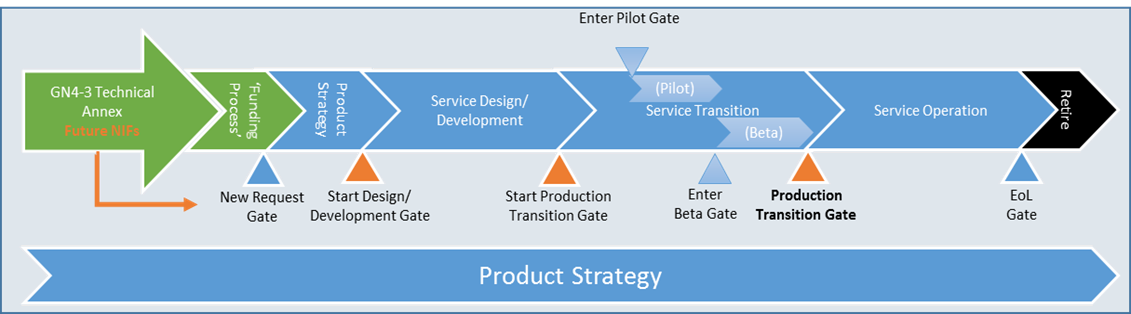}
\caption{Product Lifecycle Management process in G\'{E}ANT}
\label{fig:plm}
\end{figure}

The PLM process, depicted in Fig.~\ref{fig:plm}, is sequential and consists of multiple phases representing various stages of product development, each ending with a gate that the product needs to pass. Each of the phases is associated with various risk factors. The objective is to seek ways to mitigate or avoid these risks to maximize the chance of a successful deployment and delivery.

Benefits from the adoption of best practices for the PLM process have been found on both sides. For the software teams, they usually include the following:
\begin{enumerate}
    \item Re-using the experience of other teams, expressed by best practices.
    \item Promoting agile mindset in software development: (1) push for continual/incremental improvement: small steps, safe investment, (2) experimenting, (3) evidence-based approach.
\end{enumerate}

The perspective of the PLM team is slightly different:
\begin{enumerate}
    \item Improved coordination of efforts in various teams: good practices are appreciated, bad practices are improved or eliminated.
    \item Instead of a fixed, imposed policy, the organization focuses on continual improvement and experimenting with processes.
    \item Promotion and implementation of continual learning, by constantly refining the best practices.
    \item Products pass the PLM gates faster and more effectively.
\end{enumerate}
The process of adopting the best practices framework into the PLM process has not finished yet. However, the early results are promising and show the versatility of the framework.

\section{Summary}
\label{sec:summary}

Implementing the outcomes of process improvement initiatives is complex and comprises several stages. In the G\'{E}ANT case, it is additionally challenging by the distributed structure of the project and the independence of the software teams.
In the paper, we have presented observations from the adoption of the best practices for software teams, both from the teams' point of view, as well as from the managerial perspective. This shows the potential of the best practices framework, both as a generic tool aimed to support the software organizations in their improvement efforts, and as a set of elaborated recommendations. We show that practices need to be adjusted to the local context to match the need and setting of a specific team and recommend how they could be monitored and refined.
We plan to continue the work towards summarizing the entire SPI efforts in G\'{E}ANT and concluding the results with quantitative results.

\bibliographystyle{unsrt}
\bibliography{gsmm}

\end{document}